\documentclass[aps,reprint,twocolumn,superscriptaddress]{revtex4-2}

\usepackage{graphics} 
\usepackage{mathtools}
\usepackage{amsmath,amssymb}
\usepackage{hyperref}  
\usepackage{graphicx}
\usepackage{dsfont}
\usepackage{color}
\usepackage{setspace}
\begin{document}

\title{Kineclinic magnetogenesis in relativistic collisionless plasmas}

\author{Modhuchandra Laishram} 
\affiliation{Asia Pacific Center for Theoretical Physics, Pohang, Gyeongbuk 37673, Republic of Korea} 

\author{Suresh Basnet}
\affiliation{Asia Pacific Center for Theoretical Physics, Pohang, Gyeongbuk 37673, Republic of Korea}

\author{Young Dae Yoon}\email{youngdae.yoon@apctp.org}
\affiliation{Asia Pacific Center for Theoretical Physics, Pohang, Gyeongbuk 37673, Republic of Korea}
\affiliation{Department of Physics, Pohang University of Science and Technology, Pohang, Gyeongbuk 37673,
Republic of Korea}

\date{\today}

\begin{abstract}
The relativistic momentum equation of a collisionless plasma is reformulated to describe the time evolution of canonical vorticity. Compared to the non-relativistic counterpart, an additional source term for canonical vorticity is identified, which embodies the misalignment between the fluid momentum and fluid velocity gradients. This kineclinic term breaks the frozen-in condition of canonical vorticity, thereby enabling generation or dissipation of magnetic fields and vorticity. We verify the role of this effect through particle-in-cell simulations of a modified Beltrami flow. Kineclinicity should be finite for all relativistic plasma systems due to the general lack of a functional relationship between fluid momentum and fluid velocity. 
\end{abstract}
\maketitle
The dynamical behavior of astrophysical plasmas spans a broad spectrum of regimes. While jets in planetary nebulae and protostellar systems are typically non-relativistic, those in active galactic nuclei (AGN) and gamma ray bursts (GRBs) are relativistic or even ultra-relativistic~\cite{Anglada2018, Blandford2019, Joshi2023}. In addition, many high-energy environments such as pulsar wind nebula (PWN) and supernova remnants (SNRs) are typically relativistic pair plasmas with Lorentz factors reaching up to 100~\cite{Liang2013, Nishikawa2013,Nishikawa2014,Chen2023}. Magnetic fields of varying scales and magnitudes are ubiquitous in these extreme plasma systems, and they are responsible for numerous dynamical processes~\cite{Parker1979, Bonafede2010,Takabe2021, Brandenburg2005, Vachaspati2021}. 
The origin of the magnetic fields and their amplification process -- magnetogenesis -- of these extreme plasmas remains an interdisciplinary subject of serious debate 
from both cosmological \cite{Vachaspati2021,Jedamzik2020}, and plasma 
physics aspects~\cite{Brandenburg2012, Zhou2022}, and both perspectives are justified in 
different astrophysical scenarios~\cite{Widrow2002,Garaldi2021}.  

It is often thought that large-scale magnetic fields originate from the amplification of a microscopic “seed field” via a dynamo process~\cite{Brandenburg2005,Brandenburg2012,Mason2011, Zhou2024, Grassi2017}. 
Popular plasma physics mechanisms that generate these seed fields include the Biermann battery (baroclinicity) effect \cite{Biermann1950},              
the Weibel instability~\cite{Weibel1959}, and 
electron scale Kelvin–Helmholtz instability(ESKHI)~\cite{gruzinov2008, Alves2012, Grismayer2013}.  

Recently, \citet{Laishram2024} showed that by taking canonical vorticity---the curl of the canonical momentum---as the primary variable, one can reformulate the momentum equation into an induction equation, akin to the plasma induction equation in magnetohydrodynamics (MHD). This formulation allows for a clear distinction and unification of various magnetogenesis mechanisms. In particular, a non-ideal term due to the pressure tensor gives rise to a canonical vorticity source that is the ``canonical battery'' effect \cite{ Yoon2019a,Laishram2024}, which generalizes both Biermann battery and Weibel instability and also predicts new mechanisms. The formulation is also useful when analyzing magnetic reconnection and helicity transport \cite{Yoon2017, Yoon2018, Yoon2019a,You2016, VonDerLinden2018}.

One limitation of the previous formulation is that it was derived for sub-relativistic velocities.  A natural question to ask is whether a similar formulation is possible in relativistic regimes. In fact, many previous works studying Weibel instability and ESKHI consider relativistic (pair) plasmas \cite{Chen2023,Liang2013, Nishikawa2013,Nishikawa2014,Nerush2017,Kagan2015}.
There is also significant effort on covariant formulations of canonical vorticity and helicity, where twisting of spacetime can give rise to magnetogenesis. ~\cite{Mahajan2003,Mahajan2010,Kawazura2014}. This mechanism arises within relativistic ideal-magnetohydrodynamics and corresponds to an additional relativistic baroclinic term. As elegant as this covariant formulation is, it makes several important assumptions, such as perfect fluid and local Maxwellian closures, that make it inaccurate for kinetic, collisionless systems. It is also difficult to compare to lab-frame experiments or simulations because magnetic field and vorticity are defined in synchronic space: a sentiment shared by some previous work \cite{Taub1948,Wright1975,Hesse2007}.

In this Letter, we formulate the canonical vorticity framework in the relativistic regime. Taking the first moment of the relativistic Vlasov equation and taking the curl, we obtain an equation describing the time evolution of canonical vorticity. Compared to the non-relativistic case, an additional term arises due to the misalignment of the fluid momentum and velocity gradients, or what we call ``kineclinicity.'' This term acts as a non-ideal term that destroys the conservation of canonical vorticity flux. We perform particle-in-cell simulations of a modified Beltrami flow to demonstrate kineclinicity and its effect in suppressing the induction of the magnetic field and vorticity. Various implications are discussed. 

\textit{Relativistic Canonical Vorticity Framework}---
To generalize the canonical vorticity framework to the relativistic regime, we first consider the relativistic Vlasov equation under the Lorentz force,
\begin{align}
    \frac{\partial f_\sigma}{\partial t}+\frac{\mathbf{p}}{\gamma m_\sigma}\cdot\frac{\partial  f_\sigma}{\partial\mathbf{x}}+q_\sigma\left(\mathbf{E}+\frac{\mathbf{p}}{\gamma m_\sigma}\times\mathbf{\mathbf{B}}\right)\cdot\frac{\partial f_\sigma}{\partial\mathbf{p}}=0,
    \label{eq:Vlasov}
\end{align}
where $f_\sigma=f_\sigma\left(\mathbf{x},\mathbf{p},t\right)$, $q_\sigma$, and $m_\sigma$ are the phase-space distribution function, charge, and mass of species $\sigma$, and $\gamma = \sqrt{1+p^2/m_\sigma^2 c^2}$ is the Lorentz factor. 

Taking the first moment of Eq.~(\ref{eq:Vlasov}), i.e., multiplying by $\mathbf{p}$ and integrating over all $\mathbf{p}$, yields the following momentum equation:
\begin{align}
    \frac{\partial \left< \mathbf{p} \right>_\sigma}{\partial t} + {\bf u}_\sigma\cdot \nabla \left< \mathbf{p} \right>_\sigma 
    ={q_\sigma}  ({\bf E}+ {\bf u}_\sigma\times {\bf B}) - \frac{\nabla\cdot {\tensor{\mathcal{P}}_\sigma}}{n_\sigma},
    \label{EOM}
\end{align}
where, denoting $\mathbf{v}=\mathbf{p}/\gamma m_\sigma$,  $n_\sigma =\int f_\sigma d^3\mathbf{p}$ is the density, $\mathbf{u}_\sigma=n_\sigma^{-1}\int \mathbf{v} f_\sigma d^3\mathbf{p}$ is the fluid velocity, $\left<\mathbf{p}\right>_\sigma=n_\sigma^{-1}\int \mathbf{p} f_\sigma d^3\mathbf{p}$ is the fluid momentum, and 
$ \tensor{\mathcal{P}}_\sigma = \int {\bf v'}{\bf p'} f_\sigma d^3{\bf p}$ is the pressure tensor, 
where $\mathbf{v}' =\mathbf{v}-\mathbf{u}_\sigma$ and $\mathbf{p}' =\mathbf{p}-\left<\mathbf{p}\right>_\sigma$
are the random part of $\mathbf{v}_\sigma$ and $\mathbf{p}_\sigma$, respectively. 
For non-relativistic distributions, $\left<\mathbf{p}\right>_\sigma = m_\sigma \mathbf{u}_\sigma$ and so Eq.~(\ref{EOM}) can be re-written in terms of $\mathbf{u}_\sigma$. On the other hand, for relativistic distributions, $\left<\mathbf{p}\right>_\sigma$ is in general not a function of $\mathbf{u}_\sigma$, so both variables need to be retained~\cite{Melzani2013}.

Taking the curl of Eq.~(\ref{EOM}), defining the 
momentum vorticity $\mathbf{\Omega}_\sigma=\nabla\times\left<\mathbf{p}\right>_\sigma$, using Faraday's law $\nabla\times\mathbf{E}=-\partial\mathbf{B}/\partial t$, and using the vector identity $\nabla\times\left(\mathbf{A}\cdot\nabla\mathbf{B}+\mathbf{A}\times\nabla\times\mathbf{B}+\nabla_\mathbf{A}\left[\mathbf{A}\cdot\mathbf{B}\right]\right)=0$  in Feynman subscript notation, yield

\begin{align}\nonumber
\left[\frac {\partial{\bf\Omega}_\sigma}{\partial t} - \nabla\times \left({\bf u}_\sigma\times  {\bf\Omega}_\sigma\right)\right]
=-{q_\sigma} \left[\frac{\partial{\bf B}}{\partial t} - \nabla\times ({\bf u}_\sigma\times {\bf B})\right]  \\
 - \nabla\times\left(\frac{\nabla\cdot \tensor{\mathcal{P}}_\sigma}{n_\sigma}\right)
 +\nabla\times \nabla_{\mathbf{u}_\sigma}\left(\left<\mathbf{p}\right>_\sigma \cdot \mathbf{u}_\sigma\right).
\label{kinetic_curl_eqn}
\end{align}

Now we define the fluid canonical vorticity ${\bf Q}_\sigma~=~ {{\bf\Omega}_\sigma+{q_\sigma}{\bf B}} $ which is 
the curl of the fluid canonical momentum $\mathbf{P}_\sigma= \left<\mathbf{p}\right>_\sigma +{q_\sigma}\mathbf{A}$ where $\mathbf{A}$ is the magnetic vector potential. ${\bf Q}_\sigma$ is now the relativistic generalization of its non-relativistic counterpart in Ref. \cite{Laishram2024}. Collecting the temporal and spatial derivatives and simplifying the last term in Eq. \ref{kinetic_curl_eqn}, we obtain
\begin{align}\nonumber
\frac {\partial {\bf Q}}{\partial t} = \underbrace{\nabla\times \left({\bf u}\times {\bf Q} \right)}_{\vec{\mathcal{C}}}
\overbrace{-\nabla\times\left(\frac{\nabla\cdot \tensor{\mathcal{P}}}{n}\right)}^{\vec{\mathcal{B}}} \\
+\underbrace{\nabla\left<p\right>_{\alpha}\times\nabla u_{\alpha}}_{\vec{\mathcal{R}}},
\label{canonical_vorticity_eqn} 
\end{align}
where $\sigma$ was dropped to reduce notational clutter, and the index $\alpha$ in the last term is summed over the three spatial dimensions 
(i.e., $\mathcal{R}_k=\epsilon_{ijk}\partial_i {\left<p\right>}_{\alpha} \partial_j u_{\alpha}$ in index notation). 

Now Eq.~(\ref{canonical_vorticity_eqn}) is in the form of the canonical induction equation~\cite{Yoon2019a,Laishram2024} that describes the time evolution of $\mathbf{Q}$ with just three terms on the right-hand side. The first term is the convective term $\vec{\mathcal{C}}$, which is isomorphic to the magnetic induction term in ideal MHD and thus signifies that $\mathbf{Q}$ is frozen into and evolves with plasma's $\mathbf{u}$. Equivalently, it signifies that, ideally, $\mathbf{Q}$ flux is conserved in the fluid frame. The second term is the relativistic canonical battery effect due to the pressure tensor $\vec{\mathcal{B}}$. The third term is the relativistic kineclinic term $\vec{\mathcal{R}}$, which is our principal result. It signifies the misalignment between the fluid momentum and fluid velocity, and reduces to zero in non-relativistic regimes. Mathematically, $\mathcal{R}_k$ is the sum of 
Jacobian determinants (or Poisson brackets) $\{\left<{p}\right>_{a}, u_{a}\}_{ij}$. Eq.~(\ref{canonical_vorticity_eqn}) reveals that $\vec{\mathcal{R}}$ is an additional source/sink of $\mathbf{Q}$ affecting its frozen-in property. 

The drawback of Eq.~(\ref{canonical_vorticity_eqn}) is that it is not in a covariant form and thus is frame-dependent. However,as aforementioned, it provides a better link to lab-frame experiments and simulations because the magnetic field and vorticity are both defined in synchronic space at a certain time. Also, Eq.~(\ref{canonical_vorticity_eqn}) was derived without simplifying approximations such as scalar temperature or enthalpy.

To understand the role of ${\mathcal{\vec{R}}_{}}$ quantitatively, let us expand the convective term:
\begin{align}
	\mathcal{C}_k&=\epsilon_{ijk}\left( \partial_iu_\alpha\partial_j\left<p\right>_\alpha-\partial_i u_\alpha \partial_\alpha\left<p\right>_j - u_\alpha\partial_\alpha\partial_i\left<p\right>_j \right),\\
	&= -\mathcal{R}_k + \epsilon_{ijk}\left(-\partial_i u_\alpha \partial_\alpha\left<p\right>_j - u_\alpha\partial_\alpha\partial_i\left<p\right>_j \right),\label{eq:R_contribution_to_C}
\end{align}
which shows that $\vec{\mathcal{R}}$ in fact partially cancels $\vec{\mathcal{C}}$, which embodies the frozen-in property of $\mathbf{Q}$. Therefore, $\vec{\mathcal{R}}$ acts as a relativistic inertial term that prevents perfect convection of $\mathbf{Q}$ in relativistic regimes. 

Aside from the role of $\vec{\mathcal{R}}$, let us note the relationship between $\vec{\mathcal{B}}$ and $\vec{\mathcal{C}}$ as well. When constructing the relativistic pressure tensor, the symmetry of the momentum flux density $\mathcal{T}_{ij}=\int p_i p_jf d^3\mathbf{p}/\gamma$ was destroyed, i.e., $\mathcal{P}_{ij}=\mathcal{T}_{ij}-nu_i \left<p\right>_j$. The asymmetric part of the canonical battery $\vec{\mathcal{B}}$ is
\begin{align}
    \epsilon_{ijk}\partial_i\left[\frac{1}{n}\left<p\right>_j\partial_\alpha\left(nu_\alpha\right)+u_\alpha\partial_\alpha\left<p\right>_j\right],
\end{align}
where the second term exactly cancels the terms other than $\mathcal{R}_k$ in Eq.~(\ref{eq:R_contribution_to_C}). The first term represents compressibility effects. 

Although $\vec{\mathcal{R}}$ and parts of $\vec{\mathcal{B}}$ together cancel out $\vec{\mathcal{C}}$, the representation in Eq.~(\ref{canonical_vorticity_eqn}) explicitly decomposes the time evolution of $\mathbf{Q}$ into three distinct, physically interpretable sources. Now we shall demonstrate the role of each of these effects with the aid of particle-in-cell (PIC) simulations. 

\begin{figure*}
\includegraphics[width=\textwidth]{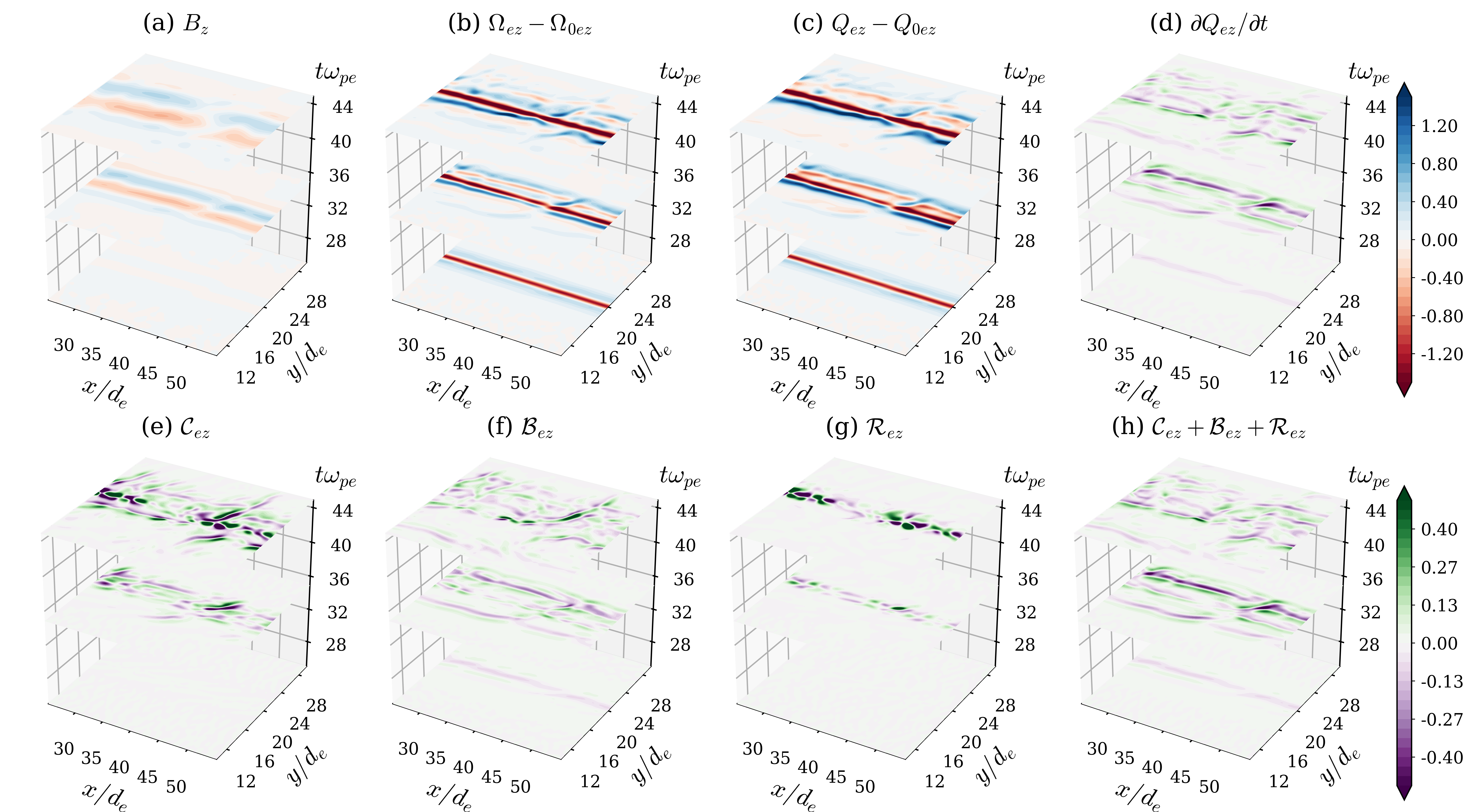}\\
\caption{Streak plots of various quantities 
(a) $B_z[m_e\omega_{pe}/e]$, 
(b) $(\Omega_{ez}-\Omega_{0ez})[m_e\omega_{pe}]$, 
(c) $(Q_{ez}-Q_{0ez})[m_e\omega_{pe}]$, 
(d) $\partial Q_{ez}/\partial t [m_e\omega_{pe}^2]$,
(e) $\mathcal{C}_z~[m_e\omega_{pe}^2]$,  
(f) $\mathcal{B}_z~[m_e\omega_{pe}^2]$, 
(g) $\mathcal{R}_{ez} ~[m_e\omega_{pe}^2]$, and 
(h) ($\mathcal{C}_{ez}+\mathcal{B}_{ez}+\mathcal{R}_{ez})~[m_e\omega_{pe}^2]$ from the 2D PIC simulation. The three slices in each panel correspond to $t\omega_{pe}=25,35,$ and $45$.}
\label{fig:2D_streak}
\end{figure*}

\textit{Numerical Results}---
To validate the model and visualize the role of each term in Eq.~(\ref{canonical_vorticity_eqn}), several 2D3V particle-in-cell (PIC) simulations were conducted for a modified relativistic Beltrami flow of a pair plasma using the SMILEI code~\cite{DEROUILLAT2018351}. The fiducial simulation domain 
was $(L_x, L_y)=(80,80)~d_e$ divided into 2048 cells in each direction, where $d_e=c/\omega_{pe}$ is the collisionless electron skin depth,
and $400$ particles were placed per cell. The spatial and temporal step sizes were $\Delta x = \Delta y =0.039d_e $, $\Delta t =0.026/\omega_{pe}$, respectively. The initial fluid velocity profile for both electrons and positrons were
$u_x=-u_{x0}[\tanh\left(\left\{y-0.25L_y\right\}/\delta\right)-\tanh\left(\left\{y-0.75L_y\right\}/\delta\right)-1], 
u_y=0$,
and $u_z=u_{z0}[\textrm{sech}\left(\left\{y-0.25L_y\right\}/\delta\right)-\textrm{sech}\left(\left\{y-0.75L_y\right\}/\delta\right)]$,
with $\delta=1d_e$, a uniform initial density $n_0$ and thermal velocity $v_{th}/c=0.097$, and periodic boundary conditions were used. 

For $u_{x0}=u_{z0}$, this flow satisfies the Beltrami flow condition $\nabla\times\mathbf{u}=\lambda\mathbf{u}$, where $\lambda$ is a scalar function \cite{Mahajan1998}. Such a system is susceptible to both ESKHI and Weibel instability, depending on the underlying parameters. However, since the focus is not on the specific details of a particular instability, we seek initial conditions for which all the terms in Eq.~(\ref{canonical_vorticity_eqn}) are roughly equally important. To satisfy this condition, the fiducial run was empirically chosen to have $v_{z0}=c\sqrt{1-\gamma^{-2}}$ where $\gamma=50$, and $v_{x0}=v_{z0}/1.5$. This condition may be relevant for, e.g., a relativistic jet along the $z$-direction with a finite hydrodynamical helicity density.

For detailed analysis, we focus on the electron dynamics at a particular shear region at $y=L_y/4$, which is similar to the mirror reflection of the shear region at $y=3L_y/4$. Fig.~\ref{fig:2D_streak} shows streak plots of various quantities from the fiducial simulation. Note that all quantities have been Gaussian filtered by 10 grid points to reduce the noise arising from taking multiple gradients. A current-sheet like $B_z$ structure (Fig.~\ref{fig:2D_streak}(a)) is self-generated along with a relatively strong $\Omega_{ez}$  (Fig.~\ref{fig:2D_streak}(b)), which together yield $Q_{ez}$ (Fig.~\ref{fig:2D_streak}(c)). Fig.~\ref{fig:2D_streak}(d) shows $\partial Q_{ez}/\partial t$, calculated with a backward difference scheme. This corresponds to the left-hand-side of Eq.~(\ref{canonical_vorticity_eqn}).

Figs.~\ref{fig:2D_streak}(e-h) show the streak plot of various terms on the right-hand-side of Eq.(\ref{canonical_vorticity_eqn}) and their total sum. The strong resemblance between Fig.~\ref{fig:2D_streak} (d) and (h) validates the canonical vorticity formulation given in Eq.~(\ref{canonical_vorticity_eqn}). The slight difference is due to numerical errors from the backward time-difference scheme used when calculating $\partial Q_{ez}/\partial t$ and usage of a spatial noise filter. The contrast between Fig.~\ref{fig:2D_streak}(e) and Fig.~\ref{fig:2D_streak}(h) indicates that the frozen-in property of $Q_{ez}$ is significantly violated during the evolution. Comparing Fig.~\ref{fig:2D_streak}(e) to Fig.~\ref{fig:2D_streak}(g) near the inflection point of $u_{x}$ ($y=20d_e$), $\mathcal{R}_{ez}$ roughly cancels out $\mathcal{C}_{ez}$, whereas $\mathcal{B}_{ez}$ is relatively weak within this region. 

\begin{figure}
\includegraphics[width=0.5\textwidth]{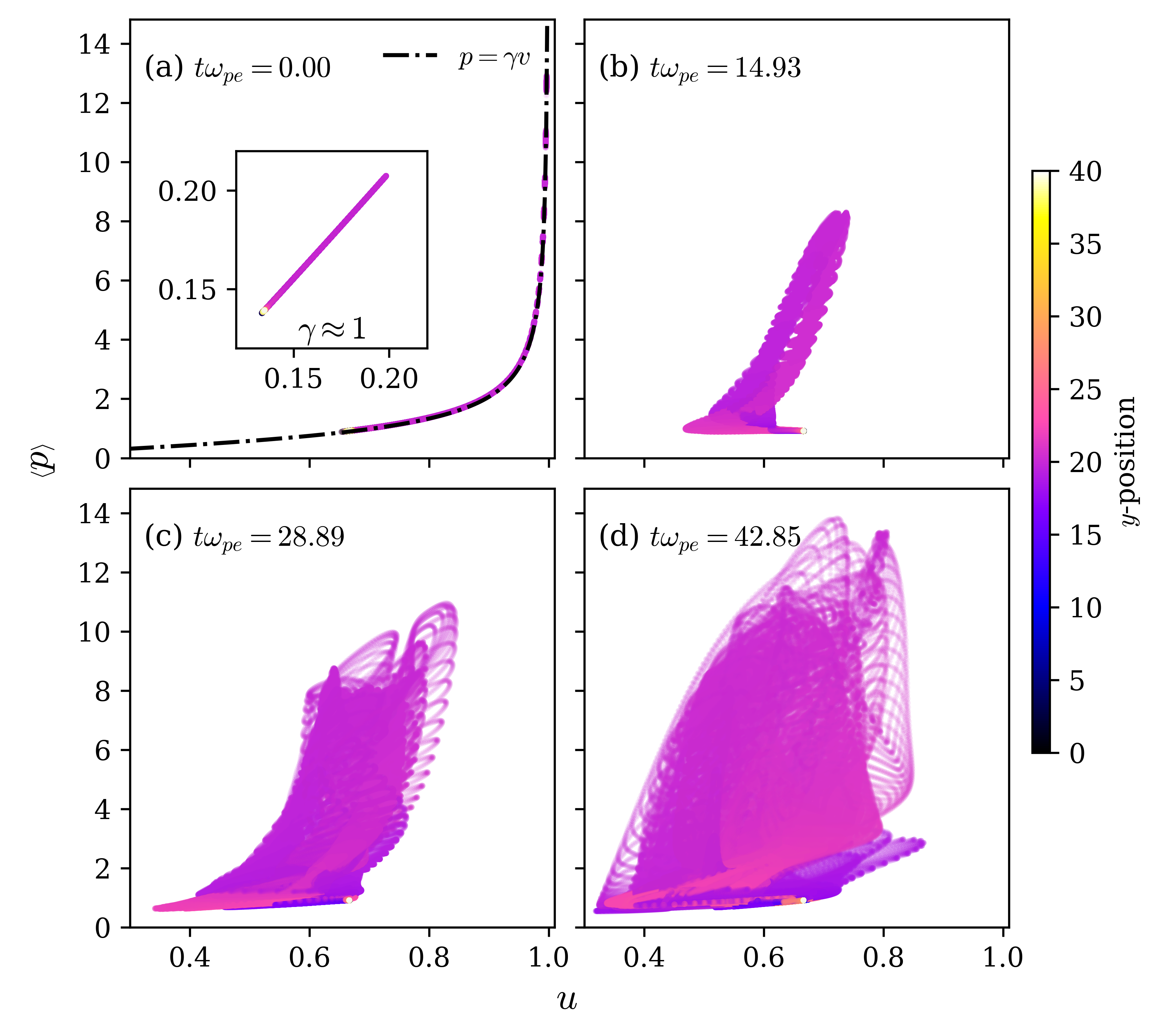}
\caption{Scatter plot of normalized $\left<{p} \right>$ vs. $u$ for 
the simulation shown in Fig.~\ref{fig:2D_streak}, displayed at times at $t\omega_{pe}\approx$ 0, 15, 30, and 45. Each scatter point corresponds to the value at a particular grid point. 
The black dashed line in panel (a) corresponds to the analytical relation 
 ${\left<p\right>}= u/\sqrt{1-u^2}$.
The inset plot in panel (a) corresponds to the same scatter plot 
for the non-relativistic $\gamma\approx 1$ case, which maintains a linear relationship at all $t\omega_{pe}$.}
\label{fig:2D_time_plot_UP_scattering}
\end{figure}

As discussed above, the kineclinic term is zero if and only if the fluid momentum $\left<\mathbf{p}\right>$ is a function of the fluid velocity $\mathbf{u}$, which is generally not true and only approximately true for non-relativistic regimes \cite{Anile_1990}. As an explicit example, if $\left<p\right>_\alpha=\gamma u_\alpha$ is assumed where $\gamma=\left(1-u_\beta u_\beta \right)^{-1/2}$ in units of momentum $mc$ and velocity $c$, we have $\partial_i \gamma = \gamma^3\, u_\beta\, \partial_i u_\beta$ and $\partial_i \langle p\rangle_\alpha
=  u_\alpha\partial_i \gamma\ + \gamma \partial_i u_\alpha 
= \gamma^3u_\alpha u_\beta\partial_i  u_\beta + \gamma \partial_i u_\alpha.$
Therefore,
\begin{align}\nonumber
\mathcal{R}_k&=\epsilon_{ijk}\partial_i {\left<p\right>_\alpha} \partial_j u_{\alpha}\\ \nonumber
&=\epsilon_{ijk}\left[ \gamma^3  u_\alpha u_\beta \partial_i u_\beta + \gamma \partial_i u_\alpha \right]\partial_j u_{\alpha}\\ \nonumber
&=\gamma^3\epsilon_{ijk}\ (u_\beta\partial_i u_\beta) (u_\alpha\partial_j u_{\alpha})
+ \gamma\epsilon_{ijk} (\partial_i u_\alpha)(\partial_j u_{\alpha})\\ \nonumber
&=0,
\end{align}
using the cyclic property of $\epsilon_{ijk}$. This is of course in general true for any $\left<p_\alpha\right>=f(u_\alpha)$ for any arbitrary function $f$ because $\partial_i \left<p\right>_\alpha=f'\partial_i u_\alpha$. Otherwise, $\vec{\mathcal{R}}$ is always finite.

Fig.~\ref{fig:2D_time_plot_UP_scattering} 
explicitly illustrates the origin of finite $\vec{\mathcal{R}}$ by showing the temporal evolution of normalized $\left<{p} \right>$ vs. $u$ for the simulation presented in Fig.~\ref{fig:2D_streak}. Each scatter point corresponds to the value of an individual grid point in the simulation. The color bar indicates the $y$-location of each fluid element across the shear interface.  Initially, the fluid momentum follows the single-particle relation $\left<p\right>=\gamma u$ due to the initial simulation setup. However, this relationship is quickly destroyed, and any functional relationship is lost, giving rise to $\vec{\mathcal{R}}$. In contrast, the inset in Fig.~\ref{fig:2D_time_plot_UP_scattering}(a) corresponds another simulation where $u_{x0},u_{z0}\ll c$. There, $u$ and 
$\left<p\right>$ are linearly correlated at all times, and so there is no kineclinicity.

Another point to note is that the fluid momentum magnitude can be much larger than unity even when the fluid velocity is non-relativistic, because the former is dominated by high-momentum population, whereas the latter is dominated by the statistical average velocity \cite{Hesse2007}. For example, consider the distribution function $f\sim\left[a\delta\left(\mathbf{p}-p_a\hat{y}\right)+b\delta\left(\mathbf{p}-p_b\hat{x}\right)\right]$ where $\delta$ is the Dirac delta function, $a$ represents a diffuse, highly relativistic population, and $b$ represents a dense, non-relativistic population. The first moments of this distribution function are, in normalized units, $\left<\mathbf{p}\right>\sim ap_a\hat{y} + bp_b\hat{x}$ and $\mathbf{u}\sim ap_a/\sqrt{1+p_a^2}\hat{y}+bp_b/\sqrt{1+p_b^2}\hat{x}$. It is trivial to see that when $a\ll b$, $p_a/p_b\gg b/a$, and $p_b$ is of order unity, it follows that $\left<\mathbf{p}\right>\simeq ap_a\hat{y}$ but $\mathbf{u}\simeq bp_b/\sqrt{1+p_b^2}\hat{x}$. This kind of difference can be seen in Fig. \ref{fig:2D_time_plot_UP_scattering}(d) where the maximum normalized fluid momentum is around 13 whereas the maximum fluid speed is only around 0.8. This is because even though most of the initially relativistic population slows down due to the instability and dominates the fluid velocity, the remaining energetic population is sufficient to maintain a large fluid momentum.

\textit{Discussion}--- 
Although $\vec{\mathcal{R}}$ partially cancels out $\vec{\mathcal{C}}$, this does not necessarily mean that kineclinicity always suppresses the generation of canonical vorticity. For instance, if the two terms in Eq.~(\ref{eq:R_contribution_to_C}) are both large but cancel each other out, $\vec{\mathcal{C}}$ is small, but $\vec{\mathcal{R}}$ can be big. This means that there may be situations where kineclinicity is dominantly responsible for canonical vorticity generation. We have attempted to isolate such a configuration but to no avail so far, so it remains to be seen whether a kineclinicity-dominated magnetogenesis can occur. 

Moreover, although only a modified Beltrami flow was simulated in this work to distinctly demonstrate the role of each term in Eq.~(\ref{canonical_vorticity_eqn}), the generality of this formulation calls for an investigation of the role of kineclinicity in various other relativistic situations. Kineclinicity may be an important fundamental feature of many relativistic scenarios, including astrophysical jets, collisionless shocks, relativistic magnetic reconnection, and many more. In particular, our preliminary study shows that the effect may be significant in relativistic magnetic reconnection, where the kineclinic term may be an additional source of topology change of $\mathbf{Q}$ and thus $\mathbf{B}$ field lines. 

The simulation in the present work involves pair plasmas to avoid the multiscale complexity and computational stiffness associated with the $m_i/m_e$ ratio. In electron–ion plasmas, one would expect substantially richer structure formation and additional physics of instabilities~\cite{ Grismayer2013, Nishikawa2014}. Because Eq.~(\ref{canonical_vorticity_eqn}) applies to any species, an ion-driven kineclinic effect should occur in analogy with the electron contribution, but on distinct timescales; determining when and to what extent this ion effect becomes important is an interesting direction for future investigation. 

\textit{Summary}--- 
In summary, the curl of the relativistic momentum equation for a collisionless plasma is reformulated in terms of the time evolution of the canonical vorticity. In comparison to the non-relativistic counterpart, the reformulation yields a term that represents the misalignment between fluid momentum and velocity gradients, which we name ``kineclinicity." This term can spontaneously generate vorticity and magnetic fields, and is a generic feature of all relativistic plasma flows due to the general absence of a functional relationship between the fluid momentum and velocity. We verify this mechanism in relation to the other two magnetogenesis sources in the formulation with the aid of particle-in-cell simulations of a modified Beltrami flow. 

\begin{acknowledgments}
This work was supported by an appointment to the JRG Program at the APCTP through the Science and Technology Promotion Fund and Lottery Fund of the Korean government, and also by the Korean Local Governments---Gyeongsangbuk-do Province and Pohang City. This work was also supported by the NRF of Korea under Grant No. NRF-RS-2023-00281272 and NRF-RS-2025-00522068. The computations presented here were conducted on the KAIROS supercomputing cluster at the Korean Institute of Fusion Energy and on the APCTP computing server.
\end{acknowledgments}

\bibliography{export1}

\end{document}